\documentclass{amsart}
\usepackage{graphicx}

\input{tcilatex}
\input tcilatex

\begin{document}
\title[superexponential population growth]{Deterministic versus stochastic aspects of superexponential population
growth models}
\author{Nicolas Grosjean and Thierry Huillet}
\address{Laboratoire de Physique Th\'{e}orique et Mod\'{e}lisation, CNRS, UMR-8089
and University of Cergy-Pontoise\\
2, rue Adolphe Chauvin F-95302, Cergy-Pontoise, Cedex, FRANCE\\
E-mail(s): nicolas.grosjean@u-cergy.fr, huillet@u-cergy.fr}
\maketitle

\begin{abstract}
Deterministic population growth models with power-law rates can exhibit a
large variety of growth behaviors, ranging from algebraic, exponential to
hyperexponential (finite time explosion). In this setup, selfsimilarity
considerations play a key role, together with two time substitutions. Two
stochastic versions of such models are investigated, showing a much richer
variety of behaviors. One is the Lamperti construction of selfsimilar
positive stochastic processes based on the exponentiation of spectrally
positive processes, followed by an appropriate time change. The other one is
based on stable continuous-state branching processes, given by another
Lamperti time substitution applied to stable spectrally positive processes.%
\newline

\end{abstract}

\section{Introduction}

Deterministic population growth models (\ref{PG1}) with power-law rates $\mu
x^{\gamma }$, $\mu >0$, can exhibit a large variety of behaviors, ranging
from algebraic ($\gamma <1$), exponential ($\gamma =1$) to hyperexponential
(finite time explosion if $\gamma >1$) growth for the size (or mass) $%
x\left( t\right) $ of some population at time $t\geq 0$. The exponential
(Malthusian) growth regime with $\gamma =1$ discriminates between the two
other ones and the transition at $\gamma =1$ is quite sharp. In this setup,
selfsimilarity considerations (with Hurst index $\alpha =1/\left( 1-\gamma
\right) $) play a key role, together with two time substitutions.
Log-selfsimilarity considerations can also be introduced while
exponentiating the latter model for $x\left( t\right) $. In this setup, the
discriminating process grows at double (or superexponential) speed. This
discriminating process separates two log-self-similar processes, one growing
at exp-algebraic rate and the other one blowing-up in finite time.

In this manuscript, two stochastic versions of such population growth models
with similar flavor are investigated, showing a much richer variety of
behaviors. One is the Lamperti construction of selfsimilar positive
stochastic processes based on the exponentiation of spectrally positive
processes, followed by an appropriate time change. As an example, the
Lamperti diffusion process (\ref{diff}) is studied in some detail, including
the noncritical cases. For the critical case with $\mu =0$ for instance, we
show that the transition $\gamma <1$ to $\gamma >1$ is rather smooth:
indeed, if $\gamma <1$, state $\infty $ is a natural inaccessible boundary
whereas state $0$ is exit (or absorbing) and reached eventually in finite
time. The population dies out (extinction) fast. If $\gamma =1$ (when the
discriminating critical process is geometric Brownian motion), state $\infty 
$ is an entrance state and state $0$ a natural inaccessible boundary. State $%
0$ (extinction) is reached eventually but now not in finite time. If $\gamma
>1$, state $0$ is a natural inaccessible boundary whereas state $\infty $ is
an entrance state. The process drifts to $\infty $ (explosion) but not in
finite time. Situations for which there is a finite time explosion can occur
but only in noncritical cases when $\gamma >1$ and $\mu $ exceeds some
positive threshold. In all cases, depending on $\gamma <1$ ($\gamma >1$),
such processes are stochastically selfsimilar with Hurst index $\alpha >0$ ($%
\alpha <0$).

The other one is based on continuous-state branching processes (CSBPs) $%
\mathrm{x}\left( t\right) $, as given by another Lamperti time substitution
of spectrally positive processes: in this respect, the $a-$stable Lamperti
CSBP (with $a\in \left( 1,2\right) $) and the one-sided $a-$stable CSBP
(with $a\in \left( 0,1\right) $) are investigated in some detail. Both
noncritical and critical cases are considered. The critical version of these
models are shown to exhibit self-similarity properties: the obtained Hurst
indices are $\alpha =1/\left( a-1\right) $ with range $\alpha >1$ and $%
\alpha <-1$, respectively. Taking $a\rightarrow 1^{\pm }$ yields in the
first place the deterministic Malthusian growth model: $\mathrm{x}\left(
t\right) =\mathrm{x}e^{\left( \mu \pm \kappa \right) t}$. This Malthusian
regime separates a situation for which $\mathbf{E}\left( \mathrm{x}\left(
t\right) \mid \mathrm{x}\left( t\right) >0\right) \propto t^{\alpha }$ has
superlinear algebraic growth rate (for the $a-$Lamperti model) and a
situation for which $\mathrm{x}\left( t\right) $ is not regular as it blows
up for all time $t>0$ (for the one-sided $a-$stable model). The Malthus
model is the discriminating critical process of such CSBP population growth
models and the situation looks quite similar to the deterministic setup,
although much more complex. The transition at $a=1$ is sharp. While
considering a different limiting process as $a\rightarrow 1^{\pm }$, we
obtain the Neveu CSBP model which grows a.s. at double superexponential
speed. The critical version of this process is no longer self-similar. It
plays the role of the superexponential discriminating deterministic model
separating two log-self-similar models: the exp-algebraic and the blowing-up
regimes, respectively.

\section{Deterministic population growth models}

\subsection{A class of self-similar growth models}

Let $x\left( t\right) \geq 0$ denote the size (mass) of some population at
time $t\geq 0$, with initially $x:=x\left( 0\right) \geq 0$. With $\mu ,$ $%
\gamma >0$, consider the growth dynamics 
\begin{equation}
\overset{.}{x}\left( t\right) =\mu x\left( t\right) ^{\gamma }\text{, }%
x\left( 0\right) =x,  \label{PG1}
\end{equation}
for some velocity field $v\left( x\right) =\mu x^{\gamma }$. Integrating
when $\gamma \neq 1$ (the non linear case), we get formally 
\begin{equation}
x\left( t\right) =\left( x^{1-\gamma }+\mu \left( 1-\gamma \right) t\right)
^{1/\left( 1-\gamma \right) }.  \label{PG1I}
\end{equation}

Three cases arise:

$\bullet $ $0<\gamma <1$: then $x\geq 0$ makes sense and in view of $%
1/\left( 1-\gamma \right) >1$, the growth of $x\left( t\right) $ is
algebraic at rate larger than $1$. We note that $x\left( t,x\right)
:=x\left( t\right) $ with $x\left( 0\right) =x$ obeys the selfsimilarity
property: for all $\lambda >0$, $t\geq 0$ and $x\geq 0$, $x\left( \lambda
t,\lambda ^{\alpha }x\right) =\lambda ^{\alpha }x\left( t,x\right) $, with $%
\alpha :=1/\left( 1-\gamma \right) >1$, the Hurst exponent. When $x=0$, the
dynamics has two solutions, one $x\left( t,0\right) \equiv 0$ for $t\geq 0$
and the other $x\left( t,0\right) =\left( \mu \left( 1-\gamma \right)
t\right) ^{1/\left( 1-\gamma \right) }$ because the velocity field $v$ in (%
\ref{PG1}) with $v\left( 0\right) =0$, is not Lipschitz as $x$ gets close to 
$0$, having an infinite derivative. The solution $x\left( t,0\right) =\left(
\mu \left( 1-\gamma \right) t\right) ^{1/\left( 1-\gamma \right) }$ with $%
x=0 $ reflects some spontaneous generation phenomenon: following this path,
the mass at time $t>0$ is not $0$, although initially it was.

$\bullet $ $\gamma >1$: then $x>0$ only makes sense and explosion or blow-up
of $x\left( t\right) $ occurs in finite time $t_{\text{exp}}=x^{1-\gamma
}/\left[ \mu \left( \gamma -1\right) \right] $. Up to the explosion time $t_{%
\text{exp}}$, $x\left( t\right) $ is selfsimilar with Hurst exponent $\alpha
=1/\left( 1-\gamma \right) <0$. Whenever $x\left( t\right) $ blows up in
finite time, following \cite{VG}, we shall speak of an hyperexponential
growth regime. This model was shown meaningful as a world population growth
model over the last two millenaries, \cite{VG}. There is also some recent
empirical interest into models with similar behavior in \cite{SA} , \cite{HS}
and \cite{JS}. The finite-time explosion feature, the related interpretation
problems and the previous works about this interpretation have been
emphasized in \cite{Ro}, where the author considers the technological
advance of a given market. More technically, necessary and sufficient
conditions for the existence of such a blowing up regime involving the
asymptotic form of the local series representation for the general solutions
around the singularities are given in \cite{GH}.

$\bullet $ $\gamma =1$: this is a simple special case not treated in (\ref
{PG1I}), strictly speaking. However, expanding the solution (\ref{PG1I}) in
the leading powers of $1-\gamma $\ yields consistently:  
\begin{equation}
\begin{array}{c}
x\left( t\right) =e^{\log \left( x^{1-\gamma }+\mu (1-\gamma )t\right)
/\left( 1-\gamma \right) } \\ 
=e^{\log [x^{1-\gamma }\left( 1+\mu x^{\gamma -1}(1-\gamma )t\right)
]/\left( 1-\gamma \right) }\sim xe^{(1/(1-\gamma ))\mu x^{\gamma
-1}(1-\gamma )t}\sim xe^{\mu t}.
\end{array}
\label{A1}
\end{equation}
Here $x\geq 0$ makes sense for (\ref{PG1}) with $x\left( t\right) =xe^{\mu t}
$ for $t\geq 0$ if $x\geq 0$. This is the simple Malthus growth model. The
Malthus regime with $\gamma =1$ will be called ``discriminating'' for (\ref
{PG1}), in the sense that it separates a slow algebraic growth regime ($%
\gamma <1$) and a blowing-up regime ($\gamma >1$).\newline

\emph{Remark:} $\left( i\right) $ One can extend the range of $\gamma $ as
follows: if $\gamma =0$, for all $x\geq 0$, $x\left( t\right) =x+\mu t$, a
linear growth regime. If $\gamma <0$, (\ref{PG1I}) holds for all $x\geq 0:$
because $1/\left( 1-\gamma \right) <1$ the growth of $x\left( t\right) $ is
again algebraic but now at rate smaller than $1$. When $\gamma \leq 0$, the
spontaneous generation phenomenon also holds with the velocity field itself
diverging near $x=0$ if $\gamma <0$: the solution $x\left( t\right) \equiv 0$
for $t\geq 0$ is no longer valid. For this range of $\gamma $, $x\left(
t\right) :=x\left( t,x\right) $ obeys the selfsimilarity property with Hurst
exponent $\alpha =1/\left( 1-\gamma \right) \in \left( 0,1\right] $.\newline

$\left( ii\right) $ One can also extend the range of $\mu $ as follows: if $%
\mu <0$, depending on $0<\gamma <1$ or $\gamma >1$, the process either goes
extinct in finite time $t_{\text{ext}}=x^{1-\gamma }/\left[ \mu \left(
\gamma -1\right) \right] $ or decays at algebraic rate $1/\left( 1-\gamma
\right) $ reaching $0$ in infinite time (respectively). Because growth is
our main interest, we shall avoid this case in general.

\subsection{Time-changes}

We shall consider two different kinds of time substitution which shall prove
of interest to us.

$\left( i\right) $ Consider the trivial dynamics $\overset{.}{s}\left( \tau
\right) =\mu $, with $s\left( \tau \right) =s\left( 0\right) +\mu \tau $,
for some clock-time $\tau \geq 0$. Let $y\left( \tau \right) =\exp s\left(
\tau \right) $. Then $\overset{.}{y}\left( \tau \right) =\mu y\left( \tau
\right) $, $y\left( 0\right) =\exp s\left( 0\right) >0$, with $y\left( \tau
\right) =y\left( 0\right) e^{\mu \tau }>0$. Consider the time substitution: $%
t_{\tau }=\int_{0}^{\tau }y\left( \tau ^{\prime }\right) ^{1/\alpha }d\tau
^{\prime }$. Then its inverse is $\tau _{t}=\int_{0}^{t}x\left( s\right)
^{-1/\alpha }ds$ where $x\left( t\right) =y\left( \tau _{t}\right) $. The
dynamics of $x\left( t\right) $ is 
\begin{equation}
\overset{.}{x}\left( t\right) =\overset{.}{y}\left( \tau _{t}\right) 
\overset{.}{\tau }_{t}=\mu x\left( t\right) ^{1-1/\alpha }.  \label{A2}
\end{equation}
It coincides with (\ref{PG1}) provided $\alpha =1/\left( 1-\gamma \right) $
or $\gamma =1-1/\alpha $. Thus $x\left( t\right) $ in (\ref{PG1}) is a
time-changed version of $y\left( \tau \right) =\exp s\left( \tau \right) $.

$\left( ii\right) $ If $s\left( \tau \right) >0$ for all $\tau \geq 0$
(requiring $s:=s\left( 0\right) >0$ and $\mu >0$), the process $y\left(
t\right) $ is itself a time-changed version of $s\left( \tau \right) $.
Consider indeed the time substitution: $t_{\tau }=\int_{0}^{\tau }s\left(
\tau ^{\prime }\right) ^{-1}d\tau ^{\prime }$. Then its inverse is $\tau
_{t}=\int_{0}^{t}y\left( s\right) ds$ where $y\left( t\right) =s\left( \tau
_{t}\right) $. The dynamics of $y\left( t\right) $ is 
\begin{equation}
\overset{.}{y}\left( t\right) =\overset{.}{s}\left( \tau _{t}\right) 
\overset{.}{\tau }_{t}=\mu y\left( t\right) .  \label{A3}
\end{equation}
If $s>0$, $\mu <0$, this is true only up to the time when $s\left( \tau
\right) $ first hits zero.

\subsection{Exponentiating and log-selfsimilarity}

Finally, with $\mu ,$ $\gamma >0$, consider now the dynamics 
\begin{equation}
\overset{.}{z}\left( t\right) =\mu z\left( t\right) \left( \log z\left(
t\right) \right) ^{\gamma }\text{, }z\left( 0\right) =z.  \label{PG2}
\end{equation}
Introducing $x\left( t\right) =\log z\left( t\right) $ and $x=\log z$, $%
x\left( t\right) $ obeys (\ref{PG1}). Integrating (\ref{PG2}), we get
formally if $\gamma \neq 1$%
\begin{equation}
z\left( t\right) =\exp \left( \left( \log z\right) ^{1-\gamma }+\mu \left(
1-\gamma \right) t\right) ^{1/\left( 1-\gamma \right) }.  \label{PG2I}
\end{equation}

We conclude:

$\bullet $ $0<\gamma <1$: the integrated solution makes sense only when $%
z\geq 1$ in which case the growth of $z\left( t\right) $ is exp-algebraic at
algebraic rate $1/\left( 1-\gamma \right) >1$. We note that with $z\left(
t,z\right) :=z\left( t\right) $ and $z\left( 0\right) =z$, $\log z\left(
t,z\right) :=x\left( t,x\right) $ obeys the self-similarity property with
Hurst exponent $\alpha =1/\left( 1-\gamma \right) >1$. So $z\left( t\right) $
is log-selfsimilar.

$\bullet $ $\gamma >1$: then $z>1$ only makes sense in general and explosion
or blow-up of $z\left( t\right) $ occurs in finite time $t_{\text{exp}%
}=\left( \log z\right) ^{1-\gamma }/\left[ \mu \left( \gamma -1\right)
\right] $. Up to the explosion time $t_{\text{exp}}$, $z\left( t\right) $ is
log-selfsimilar with Hurst exponent $\alpha =1/\left( 1-\gamma \right) <0$.
If $\gamma >1$ is an integer, values of $z<1$ are admissible.

$\bullet $ $\gamma =1$: then $z\geq 0$ makes sense for (\ref{PG2}) with
superexponential solution $z\left( t\right) =z^{e^{\mu t}}$ for $t\geq 0$.
If $z<1$, $z\left( t\right) $ decays at double exponential (or
superexponential) pace, whereas if $z>1$ growth occurs at superexponential
(or double exponential) pace, with $z\left( t\right) \equiv 1$ if $z=1$. $%
\gamma =1$ is discriminating for (\ref{PG2}) again separating a growth
regime at exp-algebraic rate and a blowing-up regime.

One can extend the range of $\gamma $ as follows: if $\gamma =0$, $z\left(
t\right) =ze^{\mu t}$, the Malthusian exponential growth regime. If $\gamma
<0$, (\ref{PG2I}) holds for all $z>0:$ because $1/\left( 1-\gamma \right) <1$%
, the growth of $z\left( t\right) $ is exp-algebraic with time now at
algebraic rate smaller than $1$ and $z\left( t\right) $ is log-selfsimilar
with Hurst exponent $\alpha =1/\left( 1-\gamma \right) \in \left( 0,1\right] 
$.

\section{Stochastic version of the self-similar growth process}

We now investigate a natural Markovian stochastic version of the positive
self-similar growth process which was first designed in \cite{Lam}. They are
obtained while considering in the latter construction a much richer class of
driving processes $s\left( \tau \right) $: the class of spectrally positive
processes with stationary independent increments. A different attempt to the
stochastization of the finite-time singularity effect was designed in \cite
{SH} and applied to the space-time clustering events and power law
Gutenberg-Richter distribution of earthquake energies.

\subsection{Spectrally positive process with stationary independent
increments}

We start with the construction of a spectrally positive process with
stationary independent increments, \cite{Kyp}.

Let $c$ and $b>0$ be two constants. Let $s\left( \tau \right) $ with $%
s:=s\left( 0\right) $ be a spectrally positive process with independent
increments and infinitesimal generator acting on $\phi \in C^{2}$, \cite
{Ber2}, 
\begin{equation}
\begin{array}{c}
G\phi \left( s\right) =\lim_{\tau \rightarrow 0^{+}}\frac{\mathbf{E}_{s}\phi
\left( s\left( \tau \right) \right) -\phi \left( s\right) }{\tau }= \\ 
\int_{0}^{\infty }\left( \phi \left( s+v\right) -\phi \left( s\right) -v\phi
^{\prime }\left( s\right) 1_{\left\{ v\leq 1\right\} }\right) \pi \left(
dv\right) +c\phi ^{\prime }\left( s\right) +\frac{1}{2}b^{2}\phi ^{^{\prime
\prime }}\left( s\right) .
\end{array}
\label{B1}
\end{equation}
$\pi $ is the L\'{e}vy measure of the jumps of $s\left( \tau \right) $,
whose support is restricted to the positive half-line; $\pi $ is assumed to
integrate $1\wedge v^{2}$. We also assume $\pi \left( dv\right) =\rho \left(
v\right) dv$ for some density function $\rho \left( v\right) $. $s\left(
\tau \right) $\ started at $s$\ has a drift term $c\tau $\ and a Brownian
component $w$\ with constant local standard deviation $b>0$\ and a pure
random jump measure term $N$\ with intensity $ds\cdot \pi \left( dv\right) $%
, specifically:\emph{\ } 
\begin{equation}
s\left( \tau \right) =s+c\tau +bw\left( \tau \right) +\int_{0}^{\tau
}\int_{0}^{\infty }vN\left( ds,dv\right) .  \label{B2}
\end{equation}
If $s>0$ and $c<0$, whenever $s\left( \tau \right) $ becomes negative, it
will do so while hitting the origin. Taking $\phi \left( s\right) =e^{-ps}$, 
$p\geq 0$, $\mathbf{E}_{s}\phi \left( s\left( \tau \right) \right) =\mathbf{E%
}_{s}e^{-ps\left( \tau \right) }$ is the Laplace-Stieltjes transform (LSt)
of $s\left( \tau \right) $ and 
\begin{equation}
G\phi \left( s\right) =-e^{-ps}\psi \left( p\right)   \label{B3}
\end{equation}
where 
\begin{equation}
\psi \left( p\right) =\int_{0}^{\infty }\left( 1-e^{-pv}-pv1_{\left\{ v\leq
1\right\} }\right) \pi \left( dv\right) +cp-\frac{1}{2}b^{2}p^{2}
\label{LLT}
\end{equation}
is the log-Laplace exponent of $s\left( \tau \right) $. As required
therefore for Markov processes with stationary independent increments, 
\begin{equation}
\mathbf{E}_{s}e^{-p\left( s\left( \tau \right) -s\right) }=e^{-\tau \psi
\left( p\right) }.  \label{B4}
\end{equation}
The function $\psi \left( p\right) $ is concave with $\psi ^{\prime }\left(
0\right) =\int_{1}^{\infty }v\pi \left( dv\right) +c=:\mu $.

\subsection{Exponential of the spectrally positive process}

We now turn to taking the exponential of the spectrally positive process $%
s\left( \tau \right) $.

Let $y\left( \tau \right) =\exp s\left( \tau \right) $, exponentiating $%
s\left( \tau \right) $. Then $y\left( \tau \right) $ with $y:=y\left(
0\right) $ is a multiplicative Markov process with infinitesimal generator ($%
\psi \left( y\right) =\phi \left( \log y\right) $), 
\begin{equation}
\begin{array}{c}
\widetilde{G}\psi \left( y\right) :=\lim_{\tau \rightarrow 0^{+}}\frac{%
\mathbf{E}_{y}\psi \left( y\left( \tau \right) \right) -\psi \left( y\right) 
}{\tau }= \\ 
\int_{1}^{\infty }\left( \psi \left( yu\right) -\psi \left( y\right) -y\log
u\psi ^{\prime }\left( y\right) 1_{\left\{ u\leq e\right\} }\right) 
\widetilde{\pi }\left( du\right) + \\ 
\left( \frac{1}{2}b^{2}+c\right) y\psi ^{\prime }\left( y\right) +\frac{1}{2}%
b^{2}y^{2}\psi ^{\prime \prime }\left( y\right) .
\end{array}
\label{B5}
\end{equation}
$\widetilde{\pi }$ is the L\'{e}vy measure of the jumps of $y\left( \tau
\right) $ started at $y$, supported by $\left( 1,\infty \right) $, with $%
\widetilde{\pi }\left( du\right) =u^{-1}\rho \left( \log u\right) du$, the
image measure of $\pi $ under the exponential transformation. If $\psi
\left( y\right) =y^{q}$, 
\begin{equation}
\begin{array}{c}
\widetilde{G}\psi \left( y\right) :=y^{q}\left( \int_{1}^{\infty }\left(
u^{q}-1-q\log u1_{\left\{ u\leq e\right\} }\right) u^{-1}\rho \left( \log
u\right) du+cq+\frac{1}{2}b^{2}q^{2}\right)  \\ 
=:y^{q}\xi \left( q\right) ,
\end{array}
\label{B6}
\end{equation}
leading to 
\begin{equation}
\mathbf{E}_{y}\left( \frac{y\left( \tau \right) }{y}\right) ^{q}=e^{\tau \xi
\left( q\right) }  \label{B7}
\end{equation}
for all $q:\xi \left( q\right) $ exists. The idea of a multiplicative
process is already present in \cite{AS}, \cite{LL2}. In these papers, a
non-linear version of the multiplicative model was used (with an extra
positive feedback not introduced here) to model explosive financial bubble
prices.

\emph{Examples:}

- If $\rho \equiv 0$ (no jumps for $s\left( \tau \right) $), $\widetilde{G}$
is the infinitesimal generator of the It\^{o} diffusion process ($\mu
=b^{2}/2+c$): 
\begin{equation}
dy\left( \tau \right) =\mu y\left( \tau \right) d\tau +by\left( \tau \right)
dw\left( \tau \right) =y\left( \tau \right) \left( \mu d\tau +bdw\left( \tau
\right) \right) ,  \label{Ito}
\end{equation}
with $w\left( \tau \right) $ the standard Brownian motion. We have $y\left(
\tau \right) =e^{s\left( \tau \right) }$ with $s\left( \tau \right) $
obeying: $ds\left( \tau \right) =cd\tau +bdw\left( \tau \right) $. In the
exponentiation process, we are led to a Malthus equation for $y$ with
randomized rate $\mu d\tau \rightarrow $ $\mu d\tau +bdw\left( \tau \right) $%
.

- Let $b=0$ and $\rho \left( v\right) =\kappa v^{-\left( 1+a\right) }/\Gamma
\left( -a\right) $, $\kappa >0$ and $a\in \left( 1,2\right) $. Then 
\begin{equation}
\begin{array}{c}
\widetilde{G}\psi \left( y\right) :=\frac{\kappa }{\Gamma \left( -a\right) }%
\int_{1}^{\infty }\left( \psi \left( yu\right) -\psi \left( y\right) -y\log
u\psi ^{\prime }\left( y\right) 1_{\left\{ u\leq e\right\} }\right) \\ 
u^{-1}\left( \log u\right) ^{-\left( 1+a\right) }du+cy\psi ^{\prime }\left(
y\right) .
\end{array}
\label{B8}
\end{equation}
If $\psi \left( y\right) =y^{q}$, 
\begin{equation}
\begin{array}{c}
\widetilde{G}\psi \left( y\right) :=y^{q}(\frac{\kappa }{\Gamma \left(
-a\right) }\int_{1}^{\infty }\left( u^{q}-1-q\log u1_{\left\{ u\leq
e\right\} }\right) \\ 
u^{-1}\left( \log u\right) ^{-\left( 1+a\right) }du+cq),
\end{array}
\label{B9}
\end{equation}
leading to 
\begin{equation}
\mathbf{E}_{y}\left( \frac{y\left( \tau \right) }{y}\right) ^{q}=e^{\tau \xi
\left( q\right) }  \label{B10}
\end{equation}
where 
\begin{equation}
\xi \left( q\right) =\frac{\kappa }{\Gamma \left( -a\right) }
\int_{1}^{\infty }\left( u^{q}-1-q\log u1_{\left\{ u\leq e\right\} }\right)
u^{-1}\left( \log u\right) ^{-\left( 1+a\right) }du+cq.  \label{B11}
\end{equation}

\subsection{Lamperti time substitution}

The self-similar process of interest is now a time-changed version of $%
y\left( \tau \right) $.

Following the path of $\left( i\right) $ in Subsection $2.2$, let $x\left(
t\right) =y\left( \tau _{t}\right) $ be a time-changed version of $y\left(
\tau \right) $ using the (now random) time substitution: $t_{\tau
}=\int_{0}^{\tau }y\left( \tau ^{\prime }\right) ^{1/\alpha }d\tau ^{\prime
} $ and its inverse $\tau _{t}=\int_{0}^{t}x\left( s\right) ^{-1/\alpha }ds$%
. Then $x\left( t\right) $ with $x:=x\left( 0\right) $ is a Markov process
with infinitesimal generator 
\begin{equation}
\begin{array}{c}
L\psi \left( x\right) :=\lim_{t\rightarrow 0^{+}}\frac{\mathbf{E}_{x}\psi
\left( x\left( t\right) \right) -\psi \left( x\right) }{t}=x^{-1/\alpha }%
\widetilde{G}\psi \left( x\right) = \\ 
\int_{1}^{\infty }\left( \psi \left( xu\right) -\psi \left( x\right) -x\log
u\psi ^{\prime }\left( x\right) 1_{\left\{ x\leq e\right\} }\right) 
\widetilde{\pi }_{x}\left( du\right) + \\ 
\left( \frac{1}{2}b^{2}+c\right) x^{1-1/\alpha }\psi ^{\prime }\left(
x\right) +\frac{1}{2}b^{2}x^{2-1/\alpha }\psi ^{\prime \prime }\left(
x\right) .
\end{array}
\label{B12}
\end{equation}
$\widetilde{\pi }_{x}$ is the L\'{e}vy measure of the jumps of $x\left(
t\right) $ with support $\left( 1,\infty \right) $ and with $\widetilde{\pi }%
_{x}\left( du\right) =x^{-1/\alpha }u^{-1}\rho \left( \log u\right) du$,
given $x\left( t\right) $ is in state $x$. Putting $\gamma =1-1/\alpha $ and 
$\mu =b^{2}/2+c$, the drift term is $\mu x^{\gamma }$ as in (\ref{PG1}).
Note that, depending on $c<-b^{2}/2$ ($\mu <0$) or $c>-b^{2}/2$ ($\mu >0$),
the drift term of $x\left( t\right) $ is either negative or positive.

It holds under some general conditions that for all $\lambda >0$, $t\geq 0$
and $x\geq 0$, $\left\{ x\left( \lambda t,\lambda ^{\alpha }x\right)
\right\} \overset{d}{=}\lambda ^{\alpha }\left\{ x\left( t,x\right) \right\} 
$, with $\alpha :=1/\left( 1-\gamma \right) $, \cite{Lam}. The stochastic
process $x\left( t\right) $ is selfsimilar with Hurst index $\alpha $, using
a terminology employed in \cite{Kol} and \cite{Man}.

\subsection{Examples}

We shall first study the purely diffusive case in some details.

$\left( i\right) $ If $\rho \equiv 0$ (no jumps for $x\left( t\right) $) a
stochastic extension of (\ref{PG1}) with continuous sample paths is ($\mu
=b^{2}/2+c$) 
\begin{equation}
\begin{array}{c}
dx\left( t\right) :=\mu x\left( t\right) ^{\gamma }dt+bx\left( t\right)
^{\left( 1+\gamma \right) /2}dw\left( t\right)  \\ 
=x\left( t\right) ^{\gamma }\left( \mu dt+bx\left( t\right) ^{\left(
1-\gamma \right) /2}dw\left( t\right) \right) \text{, }x\left( 0\right) =x>0.
\end{array}
\label{diff}
\end{equation}
Here $w\left( t\right) $ is the standard Brownian motion. The latter
Lamperti stochastic differential equation is a time-changed version of the It%
\^{o} diffusion (\ref{Ito}). Lamperti \cite{Lam} only considered (\ref{diff}%
) with $\gamma <1$. Such models were considered in \cite{SA}. They are in
the class of the so-called generalized CEV diffusion processes (see \cite
{LL1} to detect and analyze financial bubbles and \cite{RGA}), whose drift
and local volatility terms $f\left( x\right) :=\mu x^{\gamma }$\ and $%
g\left( x\right) :=bx^{\left( 1+\gamma \right) /2}$\ obey $f=Kgg^{\prime }$
for some constant $K=2\mu /\left( b^{2}\left( 1+\gamma \right) \right) $, a
possible signature of power-law stationary distribution, \cite{RGA}. Some
authors (see \cite{SE} for instance) considered a similar SDE but in the
sense of Stratonovitch. Although interesting, such SDEs fail to be
self-similar.

The invariant or speed measure density is $m\left( x\right) =g^{-2}\left(
x\right) \exp 2\int^{x}f/g^{2}\left( y\right) dy=b^{-2}x^{2\mu /b^{2}-\left(
1+\gamma \right) }$. The (non-decreasing) scale or harmonic function is 
\begin{equation}
\phi \left( x\right) =A+B\int^{x}dy\exp \left( -2\int^{y}f\left( z\right)
/g^{2}\left( z\right) dz\right) ,  \label{B13}
\end{equation}
for some constants $A,$ $B>0$, so with $\phi ^{\prime }\left( x\right)
=B\exp \left( -2\int^{x}f\left( y\right) /g^{2}\left( y\right) dy\right)
=Bx^{-2\mu /b^{2}}>0$. It is such that $\phi \left( x\left( t\right) \right) 
$ is a martingale as it kills the drift of (\ref{diff}).\newline

- If $\gamma <1$ ($\alpha >0$), the state $\infty $ is a natural
inaccessible boundary, by Feller classification of states, \cite{FW}. State $%
0$ is an exit (absorbing) state, a regular state or an entrance state
depending on $\mu \leq \gamma b^{2}/2$, $\gamma b^{2}/2<\mu <b^{2}/2$ and $%
\mu \geq b^{2}/2$, respectively and also by Feller classification of states.
In the first case, the first hitting time of $0$ given $x\left( 0\right)
=x>0 $, say $\tau _{x,0}$, is finite a.s.. So extinction occurs with
probability $1$ in finite time. In the second case, $x\left( t\right) $ is
self-similar$\left( \alpha \right) $, $\alpha =\left( 1-\gamma \right)
^{-1}>0$, only if state $0$ is made either purely absorbing or purely
reflecting. In the last case, if state $0$ is considered stationary, then $%
\tau _{x,0}$, is infinite a.s., \cite{Lam}: $x\left( t\right) $ drifts to $%
\infty $.

Regular and exit boundaries are accessible, while entrance and natural
boundaries are inaccessible. The diffusion process reaches a regular
boundary with positive probability and it can start afresh from it: one
needs to specify the boundary conditions at such a regular boundary point.
An exit boundary can also be reached from any starting point in $\left(
0,\infty \right) $ with positive probability but it is not possible to
restart the process from it: the process gets stuck or absorbed at it. The
process cannot reach an entrance boundary from any starting point in $\left(
0,\infty \right) $, but it is possible to restart the process at it. A
natural boundary cannot be reached in finite time and it is not allowed to
start the process from it.\newline

- If $\gamma =1$, then (\ref{diff}) is the discriminating process and it
coincides with (\ref{Ito}). The invariant or speed measure density is $%
m\left( x\right) =b^{-2}x^{2\mu /b^{2}-2}$. The derivative of the scale
function is $\phi ^{\prime }\left( x\right) =Bx^{-2\mu /b^{2}}>0$. By Feller
classification of states, the state $0$ is always a natural inaccessible
boundary. Let $\mu =b^{2}/2+c$. State $\infty $ is an entrance state or an
exit (absorbing) state depending on $\mu <b^{2}/2$ ($c<0$) and $\mu >b^{2}/2$
($c>0$), respectively. Thus 
\begin{equation}
x\left( t\right) \underset{t\rightarrow \infty }{\overset{a.s.}{\rightarrow }%
}\left\{ 
\begin{array}{c}
0,\text{ if }c<0 \\ 
\infty ,\text{ if }c>0
\end{array}
\right. .  \label{B14}
\end{equation}
If $c<0$, state $0$ (extinction) is reached eventually but it cannot be
reached in finite time, whereas if $c>0$, state $\infty $ is reached in
finite time (finite time hyperexponential blowing up). If $c=0$, both states 
$0$ and $\infty $ are natural boundaries and the process $x\left( t\right) $
is transient: it oscillates indefinitely between the two boundary states. We
have $x\left( t\right) =e^{s\left( t\right) }$ with $s\left( t\right) $
obeying: $ds\left( t\right) =cdt+bdw\left( t\right) $, $s\left( 0\right) =s$%
. Here, the discriminating process $x\left( t\right) =xe^{\left( \mu
-b^{2}/2\right) t+bw\left( t\right) }$ is just the geometric Brownian motion
with drift $c$ started at $x=e^{s}$, \cite{RS}. Therefore, the probability
density starting from $x>0$ that $x\left( t\right) $ is in state $y>0$ at
time $t>0$ is lognormal with 
\begin{equation}
p\left( x;t,y\right) =\frac{1}{by\sqrt{2\pi t}}e^{-\frac{1}{2b^{2}t}\left(
\log \left( y/x\right) -ct\right) ^{2}}.  \label{B15}
\end{equation}
The most probable state (or the mode) of $x\left( t\right) $ given $x\left(
0\right) =x$ is: $x_{*}\left( t,x\right) =x\exp \left( \left( \mu
-3b^{2}/2\right) t\right) $, the mean is $\mathbf{E}_{x}x\left( t\right)
=x\exp \left( \mu t\right) $ and the variance Var$\left( x\left( t\right)
\right) =x^{2}\left( e^{b^{2}t}-1\right) \exp \left( 2\mu t\right) \underset{%
t\rightarrow \infty }{\sim }x^{2}e^{2\left( \mu +b^{2}/2\right) t}$. With $%
\sigma \left( x\left( t\right) \right) =\sqrt{\text{Var}\left( x\left(
t\right) \right) }$ the standard deviation of $x\left( t\right) $, it holds
that $\sigma \left( x\left( t\right) \right) /\mathbf{E}x\left( t\right) 
\underset{t\rightarrow \infty }{\sim }e^{b^{2}t/2}\underset{t\rightarrow
\infty }{\rightarrow }\infty $, showing that the relative fluctuations of $%
x\left( t\right) $ are exponentially large and that no central limit theorem
for $x\left( t\right) $ is to be expected. The $q-$moments of $x\left(
t\right) $ are 
\begin{equation}
\mathbf{E}\left( x\left( t\right) ^{q}\right) =x^{q}\exp \left[ \left( q\mu
+q\left( q-1\right) b^{2}/2\right) t\right] ,\text{ }q>0,  \label{B16}
\end{equation}
with 
\begin{equation}
\mathbf{E}\left( x\left( t\right) ^{q}\right) \underset{t\rightarrow \infty 
}{\rightarrow }\left\{ 
\begin{array}{c}
0,\text{ if }q<1-\frac{2\mu }{b^{2}} \\ 
\infty ,\text{ if }q>1-\frac{2\mu }{b^{2}}
\end{array}
\right. .  \label{B17}
\end{equation}
Note that for $-b^{2}/2<c<0$ ($\mu >0$, $c<0$): $\mathbf{E}x\left( t\right) 
\overset{}{\underset{t\rightarrow \infty }{\rightarrow }}\infty $ together
with $x\left( t\right) \underset{t\rightarrow \infty }{\overset{a.s.}{%
\rightarrow }}0$. This process $x\left( t\right) $ lacks any self-similarity
property but it is log-selfsimilar because $x\left( t,x\right) =e^{s\left(
t,s\right) }$ with $s\left( t,s\right) =s\left( t\right) $ and $s\left(
0\right) =s$, a selfsimilar process with index $1/2$.\newline

- If $\gamma >1$ ($\alpha =\left( 1-\gamma \right) ^{-1}<0$), the state $0$
is a natural inaccessible boundary. State $\infty $ is an entrance state, a
regular state or an exit (or absorbing) state depending on $\mu \leq b^{2}/2$%
, $b^{2}/2<\mu <\gamma b^{2}/2$ and $\mu \geq \gamma b^{2}/2$, respectively.
In the first case, the first hitting time $\tau _{x,\infty }$ of $\infty $,
given $x\left( 0\right) =x>0$, is infinite a.s.. In the second case, $%
x\left( t\right) $ is self-similar$\left( \alpha \right) $ only if state $%
\infty $ is made either purely absorbing or purely reflecting. In the third
case, the first hitting time $\tau _{x,\infty }$ of $\infty $, given $%
x\left( 0\right) =x>0$, is finite a.s.: explosion occurs with probability $1$
in finite time (a case of hyperexponential growth). This results from the
following observation: consider the diffusion process (\ref{diff}) with $%
0<\gamma <1$. Consider the change of variables $\overline{x}\left( t\right)
:=x\left( t\right) ^{-\gamma }$ with state $0$ (respectively $\infty $)
mapped to state $\infty $ (respectively $0$). By It\^{o} calculus, 
\begin{equation}
d\overline{x}\left( t\right) =\overline{\mu }\overline{x}\left( t\right) ^{%
\overline{\gamma }}dt+\overline{b}\overline{x}\left( t\right) ^{\left( 1+%
\overline{\gamma }\right) /2}dw\left( t\right) ,\text{ }\overline{x}=x,
\label{diffb}
\end{equation}
where $\overline{\gamma }=1/\gamma >1$, $\overline{\mu }=\gamma \left(
b^{2}\left( \gamma +1\right) /2-\mu \right) $ and $\overline{b}=-b\gamma $.
The diffusion process (\ref{diffb}) is of the same form as (\ref{diff}) and
our conclusions follow the ones obtained in the case $0<\gamma <1$ and from
the facts: $\mu \geq b^{2}/2\Leftrightarrow \overline{\mu }\leq \overline{b}%
^{2}/2$ and $\mu \leq \gamma b^{2}/2\Leftrightarrow \overline{\mu }\geq 
\overline{\gamma }\overline{b}^{2}/2$.\newline

- The critical case ($\mu =0$): when $\mu =0$ (or $c=-b^{2}/2$), $x\left(
t\right) $ is a martingale so with $\mathbf{E}_{x}x\left( t\right) =x$,
constant \footnote{%
The use here of the terminology ``criticality'' refers to whether the
process will, on average, decrease $\mu <0$ (subcriticality), remain
constant $\mu =0$ (criticality) or increase $\mu >0$ (supercriticality).}.
From the previous study, if $\gamma <1$, the state $\infty $ is a natural
inaccessible boundary whereas state $0$ is exit (or absorbing) and reached
eventually in finite time. If $\gamma =1$ (the discriminating critical
process), state $\infty $ is an entrance state and state $0$ a natural
inaccessible boundary. Because $c=-b^{2}/2<0$, state $0$ (extinction) is
reached eventually but now not in finite time. If $\gamma >1$, state $0$ is
a natural inaccessible boundary whereas state $\infty $ is an entrance
state. The process drifts to $\infty $ but not in finite time.\newline

$\left( ii\right) $ Consider the following Lamperti spectrally positive
process $s\left( \tau \right) $:

Take $\pi \left( dv\right) =\rho \left( v\right) dv$ with $\rho \left(
v\right) =\kappa v^{-\left( 1+a\right) }/\Gamma \left( -a\right) $, $\kappa
>0$ and $a\in \left( 1,2\right) $. Assume $b=0$ (no Brownian component).
Then, when acting on $\phi \left( s\right) =e^{-ps}$, $p\geq 0$, the
infinitesimal generator of $s\left( \tau \right) $ reads 
\begin{equation}
G\phi \left( s\right) =e^{-ps}\left( \int_{0}^{\infty }\left(
e^{-pv}-1+pv1_{\left\{ v\leq 1\right\} }\right) \pi \left( dv\right)
-cp\right) =-e^{-ps}\psi \left( p\right) ,  \label{B18}
\end{equation}
where, for some new drift value $\mu =\frac{\kappa a}{\Gamma \left(
2-a\right) }+c$, 
\begin{equation}
\psi \left( p\right) =\frac{\kappa }{\Gamma \left( -a\right) }%
\int_{0}^{\infty }\left( 1-e^{-pv}-pv1_{\left\{ v\leq 1\right\} }\right)
v^{-\left( 1+a\right) }dv+cp=\mu p-\kappa p^{a}.  \label{B19}
\end{equation}
The factor $\frac{\kappa a}{\Gamma \left( 2-a\right) }$ in $\mu $
corresponds to an additional drift contribution arising from small jumps in
the jump part with density $\rho \left( v\right) $. Therefore 
\begin{equation}
\psi _{\tau }\left( p\right) :=-\log \mathbf{E}_{s}e^{-p\left( s\left( \tau
\right) -s\right) }=\tau \psi \left( p\right) .  \label{B20}
\end{equation}
Depending on $c<-\frac{\kappa a}{\Gamma \left( 2-a\right) }$ or $c>-\frac{%
\kappa a}{\Gamma \left( 2-a\right) }$, the global drift is either negative
or positive. If $\mu =0$ (no drift term), we shall speak of the critical
Lamperti model. The self-similar process $x\left( t\right) $ constructed as
a time-changed version of $y\left( \tau \right) =e^{s\left( \tau \right) }$,
with $s\left( \tau \right) $ the latter Lamperti process, deserves interest
but we shall not run into its detailed study.\newline

$\left( iii\right) $ The $a-$stable subordinator. The jump part of the
process $s\left( \tau \right) $ can be a subordinator, so with
non-decreasing sample paths and with bounded variations, \cite{Ber2}. Taking 
$\pi \left( dv\right) =\rho \left( v\right) dv$ with $\rho \left( v\right)
=\kappa av^{-\left( 1+a\right) }/\Gamma \left( 1-a\right) $, $\kappa >0$, $%
a\in \left( 0,1\right) $ and $b=0$, we are led to the one-sided $a-$stable
process with drift. Here therefore, $\psi \left( p\right) =\mu p+\kappa
p^{a} $, $\mu =\frac{\kappa a}{\Gamma \left( 2-a\right) }+c$. The
self-similar process $x\left( t\right) $ constructed from the latter $a-$%
stable process with drift $s\left( \tau \right) $ deserves interest but we
shall not run into its detailed study either.

\section{Growth processes as continuous-state branching processes (CSBPs)}

\subsection{Generalities on CSBPs}

Let $s\left( \tau \right) $ be the above spectrally positive L\'{e}vy
process defined by (\ref{LLT}). Following the time-change suggested in $%
\left( ii\right) $ of subsection $2.2$, consider now the new time
substitution: $t_{\tau }=\int_{0}^{\tau }s\left( \tau ^{\prime }\right)
^{-1}d\tau ^{\prime }$, defined up to the first hitting time of $0$ of $%
s\left( \tau \right) $. Then its inverse is $\tau _{t}=\int_{0}^{t}\mathrm{x}%
\left( s\right) ds$ where $\mathrm{x}\left( t\right) :=s\left( \tau
_{t}\right) =s\left( \int_{0}^{t}\mathrm{x}\left( s\right) ds\right) $.
Therefore, $\mathrm{x}\left( t\right) $ with $\mathrm{x}\left( 0\right) =%
\mathrm{x}$, solves the stochastic differential equation (SDE) 
\begin{equation}
\mathrm{x}\left( t\right) =\mathrm{x}+c\int_{0}^{t}\mathrm{x}\left( s\right)
ds+b\int_{0}^{t}\sqrt{\mathrm{x}\left( s\right) }dw\left( s\right)
+\int_{0}^{t}\int_{0}^{\infty }\int_{0}^{\mathrm{x}\left( t_{-}\right) }v%
\widetilde{N}\left( ds,dv,d\mathrm{x}\right) ,  \label{dyng}
\end{equation}
where $w=$\ $\left( w\left( t\right) ,\tau \geq 0\right) $\ is a standard
Brownian motion, $N(ds,dv,d\mathrm{x})$\ is a Poisson random measure with
intensity $ds\cdot \pi (dv)\cdot d\mathrm{x}$\ independent of $w$\ and $%
\widetilde{N}$\ is the compensated measure of $N$. And $\mathrm{x}\left(
t\right) $ is a continuous-state branching process (CSBP), \cite{Lam2},
stopped when it first hits $0$ if ever. From \cite{FL10} indeed, a CSBP can
also be defined as the unique non-negative strong solution of this SDE.
CSBPs may be viewed as properly scaled versions of the classical
integral-valued branching processes, \cite{Li}, \cite{Harris}, \cite{Ber2}.

Suppose $\mathrm{x}\left( 0\right) =\mathrm{x}=1$. Let then $\Psi _{t}\left(
p\right) :=-\log \mathbf{E}_{\mathrm{x}=1}e^{-p\mathrm{x}\left( t\right) }$,
the log-Laplace transform (LLt) of $\mathrm{x}\left( t\right) $.\ Then \cite
{Lam2}, $\Psi _{t}\left( p\right) $\ obeys 
\begin{equation}
\overset{.}{\Psi }_{t}\left( p\right) =\psi \left( \Psi _{t}\left( p\right)
\right) ,\ \Psi _{0}\left( p\right) =p,  \label{LLtsol}
\end{equation}
with $\psi $ given by (\ref{LLT}) known as the branching mechanism of $%
\mathrm{x}\left( t\right) $. We clearly have 
\begin{equation}
\Psi _{t}\left( p\right) =B^{-1}\left( t+B\left( p\right) \right) ,\text{
where }B\left( p\right) =\int^{p}\frac{dq}{\psi \left( q\right) }.
\label{LLts}
\end{equation}
Furthermore, with 
\begin{equation}
\Psi _{t,\mathrm{x}}\left( p\right) :=-\log \mathbf{E}_{\mathrm{x}}e^{-p%
\mathrm{x}\left( t\right) },\Psi _{t,\mathrm{x}}\left( p\right) =\mathrm{x}%
\Psi _{t}\left( p\right) .  \label{LLtI}
\end{equation}
Depending on $\psi ^{\prime }\left( 0^{+}\right) $ positive, zero or
negative, $\mathrm{x}\left( t\right) $ is supercritical, critical or
subcritical. In the supercritical case, $\mathrm{x}\left( t\right) $ started
at $\mathrm{x}>0$ has a positive extinction probability $\rho _{\mathrm{x},%
\text{ext}}=\rho _{\text{ext}}^{\mathrm{x}}$ with $\rho _{\text{ext}}:=\rho
_{1,\text{ext}}=\exp \left( -p_{c}\right) $ and $p_{c}$ the largest solution
to $\psi \left( p\right) =0$. If in the supercritical case $\psi \left(
p\right) \geq 0$ for all $p\geq 0$, by convention $p_{c}=\infty $ and
therefore $\rho _{\mathrm{x},\text{ext}}=0$ (a case of strict
supercriticality). In the critical and subcritical cases, $\mathrm{x}\left(
t\right) $ started at $\mathrm{x}>0$ goes extinct with probability $1$.

If $\tau _{\mathrm{x},0}=\inf \left( t>0:\mathrm{x}\left( t\right) =0\mid 
\mathrm{x}\left( 0\right) =\mathrm{x}\right) $ now denotes the time to
extinction, we have 
\begin{equation}
\mathbf{P}\left( \tau _{\mathrm{x},0}\leq t\right) =e^{-\mathrm{x}\Psi
_{t}\left( \infty \right) }.  \label{ExtT}
\end{equation}

\subsection{Examples}

We shall consider 3 fundamental examples:\newline

$\bullet $ $\rho \equiv 0$. We are led to the Feller diffusion on $\left[
0,\infty \right) $ (compare with (\ref{Ito})): 
\begin{equation}
d\mathrm{x}\left( t\right) =c\mathrm{x}\left( t\right) dt+b\sqrt{\mathrm{x}%
\left( t\right) }dw\left( t\right) \text{, }\mathrm{x}\left( 0\right) =%
\mathrm{x}=1.  \label{dynb}
\end{equation}

We shall let $f\left( \mathrm{x}\right) =c\mathrm{x}$, the drift and $%
g\left( \mathrm{x}\right) =b\mathrm{x}^{1/2}$, the local volatility. Note $%
g\left( \mathrm{x}\right) $ is non-Lipschitz, so singular, as $x$ approaches 
$0$. The invariant or speed measure density of this diffusion process is $%
m\left( x\right) =g^{-2}\left( x\right) \exp 2\int^{x}f/g^{2}\left( y\right)
dy=b^{-2}x^{-1}e^{2cx/b^{2}}$. Its scale or harmonic function is $\phi
\left( x\right) =A+B\int^{x}dy\exp \left( -2\int^{y}f/g^{2}\left( z\right)
dz\right) $, for some constants $A,$ $B>0$, so with $\phi ^{\prime }\left(
x\right) =B\exp \left( -2\int^{x}f/g^{2}\left( y\right) dy\right)
=Be^{-2cx/b^{2}}>0$. It is such that $\phi \left( x\left( t\right) \right) $
is a martingale. By Feller classification of states, whatever the values of $%
c$, state $0$ is absorbing, whereas state $\infty $ is an inaccessible
natural boundary, \cite{FW1}.

Here $\psi \left( p\right) =cp-\frac{1}{2}b^{2}p^{2}$ and with $\Psi
_{t}\left( p\right) :=-\log \mathbf{E}_{1}e^{-p\mathrm{x}\left( t\right) }$%
,\ then $\Psi _{t}\left( p\right) $\ obeys $\overset{.}{\Psi }_{t}\left(
p\right) =\psi \left( \Psi _{t}\left( p\right) \right) $,\ $\Psi _{0}\left(
p\right) =p$, $\Psi _{t,\mathrm{x}}\left( p\right) =\mathrm{x}\Psi
_{t}\left( p\right) $. This can be solved to give 
\begin{equation}
\Psi _{t}\left( p\right) =\left\{ 
\begin{array}{c}
pe^{ct}/\left( 1+\left( b^{2}p/\left( 2c\right) \right) \left(
e^{ct}-1\right) \right) \text{ if }c\neq 0 \\ 
\left( 2p\right) /\left( 2+b^{2}tp\right) \text{ if }c=0
\end{array}
\right. .  \label{LLtsb}
\end{equation}
We note that, when $c=0$, $\Psi _{\lambda t}\left( \lambda ^{-1}p\right)
=\lambda ^{-1}\Psi _{t}\left( p\right) $, a self-similarity property. Thus, $%
\Psi _{\lambda t,\lambda \mathrm{x}}\left( \lambda ^{-1}p\right) =\Psi _{t,%
\mathrm{x}}\left( p\right) $ and, with $\mathrm{x}\left( t,\mathrm{x}\right) 
$ the solution of (\ref{dynb}) with initial condition $\mathrm{x}\left(
0\right) =\mathrm{x}$, $\mathrm{x}\left( \lambda t,\lambda \mathrm{x}\right) 
\overset{d}{=}\lambda \mathrm{x}\left( t,\mathrm{x}\right) $ \footnote{%
This property can easily be extended to all finite-dimensional distributions.%
}: the critical Feller diffusion is self-similar with index $\alpha =1$.

The case $c<0$ ($c>0$) corresponds to a subcritical (supercritical) Feller
CSBP. $c=0$ is the critical case with $\mathrm{x}\left( t\right) $ being a
martingale. We have: 
\begin{equation}
\mathbf{E}_{\mathrm{x}}\left( \mathrm{x}\left( t\right) \right) =\mathrm{x}%
\Psi _{t}^{\prime }\left( 0\right) =\left\{ 
\begin{array}{c}
\mathrm{x}e^{ct}\text{ if }c\neq 0 \\ 
\mathrm{x}\text{ if }c=0
\end{array}
\right. .  \label{meanb}
\end{equation}

- In the supercritical case with $c>0$, the extinction probability of $%
\mathrm{x}\left( t\right) $ given $\mathrm{x}\left( 0\right) =\mathrm{x}$ is 
$\rho _{\mathrm{x},\text{ext}}=\exp \left( -\mathrm{x}p_{c}\right) =\exp
\left( -2\mathrm{x}c/b^{2}\right) $ and the law of the time to extinction $%
\tau _{\mathrm{x},0}$ given $\mathrm{x}\left( 0\right) =\mathrm{x}$ is 
\begin{equation}
\mathbf{P}\left( \tau _{\mathrm{x},0}\leq t\right) =\exp -\mathrm{x}\left[
\left( b^{2}/\left( 2c\right) \right) \left( 1-e^{-ct}\right) \right] ^{-1},
\label{Ext1}
\end{equation}
with exponential tails: $e^{ct}\mathbf{P}\left( \tau _{\mathrm{x}%
,0}>t\right) \rightarrow $constant as $t\rightarrow \infty $. If $c>0$, the
law of $\tau _{\mathrm{x},0}$ has an atom at $t=\infty $ with mass $1-\exp
\left( -2\mathrm{x}c/b^{2}\right) $, corresponding to the probability that $%
\mathrm{x}\left( t\right) $ drifts to $\infty $. If the latter event occurs,
it cannot be in finite time.

- If $c\leq 0$ (sub- and critical case), $\rho _{\mathrm{x},\text{ext}}=1$
and $\mathrm{x}\left( t\right) $ hits $0$ with probability $1$ and stays
there for ever. The law of the time to extinction $\tau _{\mathrm{x},0}$
given $\mathrm{x}\left( 0\right) =\mathrm{x}$ in this case is 
\begin{equation}
\mathbf{P}\left( \tau _{\mathrm{x},0}\leq t\right) =\left\{ 
\begin{array}{c}
\exp -\mathrm{x}\left[ \left( b^{2}/\left( -2c\right) \right) \left(
e^{-ct}-1\right) \right] ^{-1}\text{ if }c<0 \\ 
\exp -2\mathrm{x}/\left( b^{2}t\right) \text{ if }c=0
\end{array}
\right. .  \label{Ext2}
\end{equation}

- In the subcritical case ($c<0$), tails are exponential: $e^{-ct}\mathbf{P}%
\left( \tau _{\mathrm{x},0}>t\right) \rightarrow $constant as $t\rightarrow
\infty $. In the critical case ($c=0$), the law of $\tau _{\mathrm{x},0}$ is
tail-equivalent to $2\mathrm{x}/\left( b^{2}t\right) $ in that $\frac{b^{2}t%
}{2\mathrm{x}}\mathbf{P}\left( \tau _{\mathrm{x},0}>t\right) \rightarrow 1$
as $t\rightarrow \infty $; thus, $\tau _{\mathrm{x},0}$ has Pareto-like
heavy tails and the time to extinction is thus much longer statistically
than when $c<0$.\newline

$\bullet $ $b=0$ and $\pi \left( dv\right) =\rho \left( v\right) dv$ with $%
\rho \left( v\right) =\kappa v^{-\left( 1+a\right) }/\Gamma \left( -a\right) 
$, $\kappa >0$ and $a\in \left( 1,2\right) $. We are then led to the
Lamperti CSBP process $\mathrm{x}\left( t\right) $, \cite{Lam2}.\ 

Let $\Psi _{t}\left( p\right) :=-\log \mathbf{E}e^{-p\mathrm{x}\left(
t\right) }$.\ Then $\Psi _{t}\left( p\right) $\ obeys $\overset{.}{\Psi }%
_{t}\left( p\right) =\psi \left( \Psi _{t}\left( p\right) \right) $,\ $\Psi
_{0}\left( p\right) =p$ where $\psi \left( p\right) =\mu p-\kappa p^{a}$, $%
\mu =\frac{\kappa a}{\Gamma \left( 2-a\right) }+c$.\ It is a CSBP, with here 
\begin{equation}
\Psi _{t}\left( p\right) =\left\{ 
\begin{array}{c}
\left( p^{1-a}e^{-\mu \left( a-1\right) t}+\left( \kappa /\mu \right) \left(
1-e^{-\mu \left( a-1\right) t}\right) \right) ^{-1/\left( a-1\right) }\text{
if }\mu \neq 0 \\ 
\left( p^{1-a}+\kappa \left( a-1\right) t\right) ^{-1/\left( a-1\right) }%
\text{ if }\mu =0
\end{array}
\right. ,  \label{LLtsL}
\end{equation}
and $\Psi _{t,\mathrm{x}}\left( p\right) =\mathrm{x}\Psi _{t}\left( p\right) 
$. Classical (i.e. discrete-space, continuous-time Bienaym\'{e}%
-Galton-Watson) branching processes displaying similar properties with
finite mean and infinite variance were considered in \cite{SS} and \cite{AGH}%
.

The case $\mu <0$ ($\mu >0$) corresponds to a subcritical (supercritical)
Lamperti CSBP. $\mu =0$ is the critical case with $\mathrm{x}\left( t\right) 
$ a martingale.

Note that $\Psi _{t}\left( p\right) \underset{p\rightarrow 0^{+}}{%
\rightarrow }0$ for all $t>0$. The Lamperti CSBP is regular or conservative
with $\mathbf{P}\left( \mathrm{x}\left( t\right) <\infty \right) =1$.

In the critical case when $\mu =0$, with $\alpha :=1/\left( a-1\right) >1$, $%
\Psi _{\lambda t}\left( \lambda ^{-\alpha }p\right) =\lambda ^{-\alpha }\Psi
_{t}\left( p\right) $, a self-similarity property. And indeed, $\Psi
_{\lambda t,\lambda ^{\alpha }\mathrm{x}}\left( \lambda ^{-\alpha }p\right)
=\Psi _{t,\mathrm{x}}\left( p\right) $, showing that $\mathrm{x}\left(
\lambda t,\lambda ^{\alpha }\mathrm{x}\right) \overset{d}{=}\lambda ^{\alpha
}\mathrm{x}\left( t,\mathrm{x}\right) $, a self-similarity property with
index $\alpha >1$ for $\mathrm{x}\left( t\right) $.

Here, $\mathrm{x}\left( t\right) $ is the jump process with drift 
\begin{equation}
d\mathrm{x}\left( t\right) =c\mathrm{x}\left( t\right) dt+\kappa \mathrm{x}%
\left( t_{-}\right) ^{1/a}ds\left( t\right) \text{, }\mathrm{x}\left(
0\right) =\mathrm{x}=1,  \label{dynL}
\end{equation}
where $s\left( t\right) $ is the driving $a-$stable spectrally positive
L\'{e}vy process ($a\in \left( 1,2\right) $), with no superposed driving
Brownian component. For this model, $\rho _{\mathrm{x},\text{ext}}=\exp
\left( -\mathrm{x}p_{c}\right) =\exp \left( -\mathrm{x}\left( \mu /\kappa
\right) ^{1/\left( a-1\right) }\right) $ in the supercritical case $\mu >0$ (%
$1$ otherwise) and 
\begin{equation}
\mathbf{P}\left( \tau _{\mathrm{x},0}\leq t\right) =e^{-\mathrm{x}\Psi
_{t}\left( \infty \right) }=\left\{ 
\begin{array}{c}
\exp -\mathrm{x}\left( \frac{\kappa }{\mu }\left( 1-e^{-\mu \left(
a-1\right) t}\right) \right) ^{-\frac{1}{a-1}}\text{ , }\mu \neq 0 \\ 
\exp -\mathrm{x}\left( \kappa \left( a-1\right) t\right) ^{-\frac{1}{a-1}}%
\text{ , }\mu =0
\end{array}
\right. .  \label{ExtL}
\end{equation}
If $c>0$, the law of $\tau _{\mathrm{x},0}$ has an atom at $t=\infty $ with
mass $1-\exp \left( -\mathrm{x}\left( \mu /\kappa \right) ^{1/\left(
a-1\right) }\right) $, the probability of explosion $\rho _{\mathrm{x},\text{%
exp}}=1-\rho _{\mathrm{x},\text{ext}}$.

In the critical case ($\mu =0$), the law of $\tau _{\mathrm{x},0}$ is
tail-equivalent to $\mathrm{x}\left( \kappa \left( a-1\right) t\right)
^{-1/\left( a-1\right) }$ as $t\rightarrow \infty $. Thus, $\tau _{\mathrm{x}%
,0}$ has power-law heavy tails and the time to extinction is thus longer
statistically than when $\mu <0$.

We can condition the critical model on non-extinction and compute\emph{\ }$%
\mathbf{E}_{1}\left( \mathrm{x}\left( t\right) \mid \mathrm{x}\left(
t\right) >0\right) $.\emph{\ }Indeed,\emph{\ }we have (\cite{RYZ}, Theorem $%
1 $), conditionally given $\mathrm{x}\left( t\right) >0$, 
\begin{equation}
\mathbf{P}_{1}\left( \tau _{1,0}>t\right) \cdot \mathrm{x}\left( t\right) 
\overset{d}{\underset{t\rightarrow \infty }{\rightarrow }}W,  \label{C1}
\end{equation}
where the random variable $W$ has LSt $\mathbf{E}\left( e^{-pW}\right)
=1-\left( 1+p^{-\left( a-1\right) }\right) ^{-1/\left( a-1\right) }$,
therefore with finite mean $1$. $\mathrm{x}\left( t\right) $ has a
quasi-stationary regime, \cite{Yag}. We have\emph{\ }$\mathbf{P}\left( \tau
_{1,0}\leq t\right) =e^{-\Psi _{t}\left( \infty \right) }$ and so $\mathbf{P}%
\left( \tau _{1,0}>t\right) =1-e^{-\Psi _{t}\left( \infty \right) }\sim \Psi
_{t}\left( \infty \right) $. This shows that as $t$ gets large 
\begin{equation}
\mathbf{E}_{1}\left( \mathrm{x}\left( t\right) \mid \mathrm{x}\left(
t\right) >0\right) \sim \frac{-1}{\Psi _{t}\left( \infty \right) }\partial
_{p}\left( \Psi _{t}\left( \infty \right) -\Psi _{t}\left( p\right) \right)
\mid _{p=0}=  \label{Yag}
\end{equation}
\begin{equation*}
\frac{1}{\Psi _{t}\left( \infty \right) }=\left( \kappa \left( a-1\right)
t\right) ^{1/\left( a-1\right) },
\end{equation*}
displaying slow algebraic superlinear growth in time, with exponent $\alpha
=1/\left( a-1\right) >1$.\newline

$\bullet $ Taking $\pi \left( dv\right) =\rho \left( v\right) dv$ with $\rho
\left( v\right) =\kappa av^{-\left( 1+a\right) }/\Gamma \left( 1-a\right) $, 
$\kappa >0$, $a\in \left( 0,1\right) $, $b=0$, we are led to the standard
one-sided $a-$stable subordinator process $s\left( \cdot \right) $ with
drift. Note that $\pi $ now integrates $1\wedge v:$ small jumps are less
likely than in the $a-$stable spectrally positive case with $a\in \left(
1,2\right) $, but large jumps of the one-sided $a-$stable subordinator are
more likely to occur than in the spectrally positive case. For this model, $%
\psi \left( p\right) =\mu p+\kappa p^{a}$, $\mu =\frac{\kappa a}{\Gamma
\left( 2-a\right) }+c$. The $\Psi _{t}\left( p\right) $ solving (\ref{LLtsol}%
) of the corresponding CSBP is seen to be 
\begin{equation}
\Psi _{t}\left( p\right) =\left\{ 
\begin{array}{c}
\left( p^{1-a}e^{\mu \left( 1-a\right) t}+\left( \kappa /\mu \right) \left(
e^{\mu \left( 1-a\right) t}-1\right) \right) ^{1/\left( 1-a\right) }\text{
if }\mu \neq 0 \\ 
\left( p^{1-a}+\kappa \left( 1-a\right) t\right) ^{1/\left( 1-a\right) }%
\text{ if }\mu =0
\end{array}
\right. ,  \label{LLtsS}
\end{equation}
and $\Psi _{t,\mathrm{x}}\left( p\right) =\mathrm{x}\Psi _{t}\left( p\right) 
$. We note that, when $\mu =0$, $\Psi _{\lambda t}\left( \lambda ^{1/\left(
1-a\right) }p\right) =\lambda ^{1/\left( 1-a\right) }\Psi _{t}\left(
p\right) $, a self-similarity property. With $\alpha =1/\left( a-1\right) $,
we have $\Psi _{\lambda t,\lambda ^{\alpha }\mathrm{x}}\left( \lambda
^{-\alpha }p\right) =\Psi _{t,\mathrm{x}}\left( p\right) $ showing that $%
\mathrm{x}\left( \lambda t,\lambda ^{\alpha }\mathrm{x}\right) \overset{d}{=}%
\lambda ^{\alpha }\mathrm{x}\left( t,\mathrm{x}\right) $, a self-similarity
property with index $\alpha <-1$. We have, 
\begin{equation}
\Psi _{t}\left( p\right) \underset{p\rightarrow 0^{+}}{\rightarrow }\left\{ 
\begin{array}{c}
\left( \left( \kappa /\mu \right) \left( e^{\mu \left( 1-a\right)
t}-1\right) \right) ^{1/\left( 1-a\right) }\text{ if }\mu \neq 0 \\ 
\left( \kappa \left( 1-a\right) t\right) ^{1/\left( 1-a\right) }\text{ if }%
\mu =0
\end{array}
\right. ,  \label{LLtsLim}
\end{equation}
and, the limit being non zero for all $t>0$, this CSBP is non-conservative
as it loses mass at $\infty $ instantaneously, with $\mathbf{P}_{\mathrm{x}%
}\left( \mathrm{x}\left( t\right) <\infty \right) =e^{-\mathrm{x}\Psi
_{t}\left( 0\right) }$. This is a consequence of $\int_{0^{+}}dq/\psi \left(
q\right) <\infty $ leading to this superexponential growth situation. Here, $%
\mathrm{x}\left( t\right) $ is the jump process with drift 
\begin{equation}
d\mathrm{x}\left( t\right) =c\mathrm{x}\left( t\right) dt+\kappa \mathrm{x}%
\left( t_{-}\right) ^{1/a}ds\left( t\right) \text{, }\mathrm{x}\left(
0\right) =\mathrm{x}=1,  \label{dynS}
\end{equation}
where $s\left( t\right) $ is the driving $a-$stable subordinator ($a\in
\left( 0,1\right) $) with no Brownian component. For this supercritical
model with $\psi ^{\prime }\left( 0^{+}\right) =\infty $, $\rho _{\mathrm{x},%
\text{ext}}=\exp \left( -\mathrm{x}p_{c}\right) =\exp \left( -\mathrm{x}%
\left( -\mu /\kappa \right) ^{1/\left( a-1\right) }\right) $ if $\mu <0$ ($0$
otherwise) and $\tau _{\mathrm{x},0}=\infty $ with probability $1$ as a
result of $B\left( p\right) =\int^{p}dq/\psi \left( q\right) \underset{%
p\rightarrow \infty }{\rightarrow }\infty $ leading to $\Psi _{t}\left(
p\right) \underset{p\rightarrow \infty }{\rightarrow }\infty $. If $\mu \geq
0$ indeed, $\psi \left( p\right) $ stays positive with $\psi \left( p\right) 
\underset{p\rightarrow \infty }{\rightarrow }\infty $ with by convention $%
p_{c}=\infty $ and so $\rho _{\mathrm{x},\text{ext}}=0$.\newline

$\bullet $ The critical growth case: the Neveu model. It remains to consider
the case $a\rightarrow 1$.

- Considering the branching mechanism of the $1-$sided $a-$stable
subordinator: $\psi \left( p\right) =\mu p+\kappa p^{a}$, $a\in \left(
0,1\right) $, $\kappa >0$ (respectively the one of the $a-$stable Lamperti
spectrally positive L\'{e}vy process: $\psi \left( p\right) =\mu p-\kappa
p^{a}$, $a\in \left( 1,2\right) $, $\kappa >0$) and letting simply $%
a\rightarrow 1^{-}$ (respectively $a\rightarrow 1^{+}$), we are led to the
branching mechanism of the pure drift model $\psi \left( p\right) =\left(
\mu +\kappa \right) p$ (respectively $\psi \left( p\right) =\left( \mu
-\kappa \right) p$). The corresponding CSBP is Malthusian and trivial: $%
\mathrm{x}\left( t,\mathrm{x}\right) =\mathrm{x}e^{\left( \mu +\kappa
\right) t}$ (respectively $\mathrm{x}\left( t,\mathrm{x}\right) =\mathrm{x}%
e^{\left( \mu -\kappa \right) t}$). This process lacks any self-similarity
property.

There is a more interesting way to take the limits $a\rightarrow 1^{\mp }:$

- Consider the branching mechanism of the $1-$sided $a-$stable subordinator: 
$\psi \left( p\right) =\mu p+\kappa p^{a}$, $a\in \left( 0,1\right) $, $%
\kappa >0$. Define the constants $\mu ^{\prime }$, $\kappa ^{\prime }>0$ by $%
\mu =\mu ^{\prime }-\kappa $ and $\kappa =\kappa ^{\prime }/\left(
1-a\right) $. Then, as $a\rightarrow 1^{-}$, $\mu \rightarrow -\infty $ and $%
\kappa \rightarrow +\infty $ in a suitable way. And $\psi $ reads $\psi
\left( p\right) =\mu ^{\prime }p-\frac{\kappa ^{\prime }}{1-a}p\left(
1-p^{a-1}\right) \sim \mu ^{\prime }p-\kappa ^{\prime }p\log p$.

- Consider the branching mechanism of the $a-$stable Lamperti spectrally
positive L\'{e}vy process: $\psi \left( p\right) =\mu p-\kappa p^{a}$, $a\in
\left( 1,2\right) $, $\kappa >0$. Define $\mu ^{\prime }$, $\kappa ^{\prime
}>0$ by $\mu =\mu ^{\prime }+\kappa $ and $\kappa =\kappa ^{\prime }/\left(
a-1\right) $. Then, as $a\rightarrow 1^{+}$, both $\mu $, $\kappa $ tend to $%
+\infty $. And $\psi $ reads $\psi \left( p\right) =\mu ^{\prime }p+\frac{%
\kappa ^{\prime }}{a-1}p\left( 1-p^{a-1}\right) \sim \mu ^{\prime }p-\kappa
^{\prime }p\log p$.

The CSBP with new branching mechanism, say $\psi \left( p\right) =\mu
p-\kappa p\log p$, $\kappa >0$, is the Neveu CSBP, \cite{Neveu}. $\mu <0$, $%
\mu =0$ and $\mu >0$ correspond respectively to the subcritical, critical
and supercritical versions of the Neveu process. Note $\psi ^{\prime }\left(
0^{+}\right) =+\infty $, so that $\mathbf{E}_{\mathrm{x}}\left( \mathrm{x}%
\left( t\right) \right) =+\infty $ and it may be shown, using martingale
arguments \cite{Neveu}, that, conditionally given $\mathrm{x}\left( t\right) 
$ drifts to $\infty $, it does so at double exponential speed a.s.. So if
the population does not go extinct, $\mathrm{x}\left( t\right) $ grows fast
to infinity at a double-exponential speed: $e^{-\kappa t}\log \mathrm{x}%
\left( t\right) ^{{}}\overset{d}{\rightarrow }E>0$ as $t\rightarrow \infty $%
, with $E$ standard exponentially distributed: $\mathbf{P}\left( E>\mathrm{x}%
\right) =e^{-\mathrm{x}}$, $\mathrm{x}>0$. Using martingale arguments, this
convergence can be shown to be almost sure as well, (\cite{Grey2}, \cite
{Neveu}, \cite{Henard}, Proposition $3.8$).

The LLt $\Psi _{t}\left( p\right) $ of the corresponding CSBP solving (\ref
{LLtsol}) is easily seen to be 
\begin{equation}
\Psi _{t}\left( p\right) =\left\{ 
\begin{array}{c}
\exp \left( \frac{\mu }{\kappa }\left( 1-e^{-\kappa t}\right) \right)
p^{e^{-\kappa t}}\text{ if }\mu \neq 0 \\ 
p^{e^{-\kappa t}}\text{ if }\mu =0
\end{array}
\right. .  \label{LLtsN}
\end{equation}
The marginal distribution of the critical Neveu CSBP is one-sided $%
e^{-\kappa t}-$stable. It holds that $\Psi _{t}\left( p\right) \underset{%
p\rightarrow 0^{+}}{\rightarrow }0$ for all $t>0$ and the critical Neveu
CSBP is regular or conservative, with $\mathbf{P}_{1}\left( \mathrm{x}\left(
t\right) <\infty \right) =1$. It can be shown that, for the critical Neveu
process, $\mathrm{x}\left( t\right) \underset{t\rightarrow \infty }{\overset{%
a.s.}{\rightarrow }}0$ (extinction a.s.), but not in finite time, \cite{H2}.
We observe that the critical version of the Neveu model lacks any
self-similarity space/time property.

\subsection{Summary}

Let us summarize our results:\newline

We considered mainly 3 fundamental CSBPs $\mathrm{x}\left( t\right) $: the
Feller diffusion model ($a=2$), the $a-$Lamperti CSBP ($a\in \left(
1,2\right) $) and the one-sided $a-$stable CSBP ($a\in \left( 0,1\right) $):

The critical version of these models were shown to exhibit self-similarity
properties: the obtained Hurst indices are $\alpha =1$, $\alpha =1/\left(
a-1\right) >1$ and $\alpha =1/\left( a-1\right) <-1$, respectively. To some
extent, the Feller diffusion model may be viewed as the limiting situation $%
a\rightarrow 2^{-}$ of the Lamperti CSBP.

Taking $a\rightarrow 1^{\mp }$ yields in the first place the deterministic
Malthusian growth models: $\mathrm{x}\left( t\right) =\mathrm{x}e^{\left(
\mu \mp \kappa \right) t}$. This Malthusian regime separates a situation for
which conditionally given $\mathrm{x}\left( t\right) >0$, the mean of $%
\mathrm{x}\left( t\right) $ has superlinear algebraic growth rate (for the $%
a-$Lamperti model, see (\ref{Yag})) and a situation for which $\mathrm{x}%
\left( t\right) $ is not regular as it blows up for all time $t>0$ (for the
one-sided $a-$stable model). It is the discriminating critical process of
such CSBP population growth models. This should be compared with similar
behaviors obtained in the deterministic setup. A main difference of the
stochastic dynamics as compared to the deterministic case is that all
critical CSBPs go extinct with probability $1$.

While considering a different limiting process as $a\rightarrow 1^{\mp }$,
we obtained the Neveu CSBP model which grows a.s. at double superexponential
speed. The critical version of this process is no longer self-similar. It
plays the role of the superexponential discriminating deterministic model
separating two log-self-similar models: the exp-algebraic rate model and the
blowing-up model, respectively.\newline

\textbf{Acknowledgments:}

T. Huillet acknowledges partial support from the ``Chaire \textit{Mod\'{e}%
lisation math\'{e}matique et biodiversit\'{e}''.} N. Grosjean and T. Huillet
also acknowledge support from the labex MME-DII Center of Excellence (%
\textit{Mod\`{e}les math\'{e}matiques et \'{e}conomiques de la dynamique, de
l'incertitude et des interactions}, ANR-11-LABX-0023-01 project). Both
authors are indebted to their referees for constructive remarks and for
bringing to their attention some important related works they were not aware
of.

\end{document}